\documentstyle[epsf,aps,prl,12pt]{revtex}
\begin{document}

\begin{center}

{\Large {\sc Turbulent--like Diffusion in Complex Quantum Systems }}

\vspace{1.8cm}

Dimitri KUSNEZOV$^a$\footnote{Internet: dimitri@nst.physics.yale.edu}, 
Aurel BULGAC$^b$\footnote{Internet: bulgac@phys.washington.edu} and
 Giu DO DANG$^c$\footnote{Internet: dodang@psisun.u-psud.fr}\\

\vspace{0.5cm}

{\sl $^a$ Center for Theoretical Physics, Sloane Physics Laboratory,\\Yale
University, New Haven, CT 06520-8120, USA}

{\sl $^b$ Department of Physics, FM--15, University of
         Washington, Seattle, WA 98195, USA}

{\sl $^c$ Laboratoire de Physique Th\'eorique et Hautes Energies,\\
      Universit\'e de Paris--Sud,  B\^at. 211, 91405 Orsay, FRANCE}

\vskip 1 cm

{\it May 1997}

\vspace{1 cm}

\parbox{13.0cm}
{\begin{center}\large\sc ABSTRACT \end{center}
{\hspace*{0.3cm} 
We study a quantum particle propagating through a ``quantum mechanically
chaotic'' background, described by parametric random matrices with only  short
range spatial correlations. The particle is found to exhibit turbulent--like
diffusion under very general situations,  without the {\it apriori}
introduction of power law noise or scaling in the background properties.}}
\end{center}


\newpage

%%%%%%%%%%%%%%%%%%%%%%%%%%%%%%%%%%%%%%%%%%%%%%%%%%%%%%%%%%%%%%%%%%%%%%%

The diffusion of particles in a complex background, such as disordered
media, has often an anomalous or L\'evy type character \cite{bou,long}.
In contrast to classical diffusion, where the variance of the
displacement of a particle or tracer  grows linearly in time, anomalous
transport can have $ \left\langle R^2\right\rangle \propto t^\alpha $,
with $\alpha \neq 1$,  although it is not limited to power law behavior.
The diffusive properties of such systems can hence range from enhanced
to dispersive dynamics. The microscopic understanding of such stochastic
processes in terms of the underlying (chaotic) dynamics remains an
active open area of study \cite{long}.  While the conventional limits of
L\'evy processes are $0<\alpha \leq 2$, there exist superdiffusive
dynamics beyond this range. One of the extreme examples is turbulent
diffusion, which occurs when the background is turbulent, originally put
in evidence in atmospheric measurements\cite{ric}. In such situations,
the average separation $R$ between two tracers can increase as fast as
$\left\langle R^2\right\rangle \sim t^3$. Richardson \cite{ric}
postulated a Fokker-Planck equation of the form:
%-----------------------------------------------------------------------
\begin{equation}
\frac{\partial {\cal P}(R,t)}{\partial t}=\frac \partial {\partial
R}\left[ V(R) \frac{\partial {\cal P}(R,t)}{\partial R}\right] ,\qquad
V(R)\propto R^{4/3},
\end{equation}
%-----------------------------------------------------------------------
which by design reproduced his measurements. Kolmogorov, in studies of
turbulence, proposed an energy--wavenumber scaling law (known as the
$\frac{5}{3}$--law) and further suggested that dissipative behavior is
spatially dependent\cite{kol}. Refinements of these scaling arguments
incorporated intermittent corrections into the power law
behavior\cite{Man}. More recently, a new class of random walks, termed
L\'evy walks, have incorporated Kolmogorov's scaling to derive
Richardson's $t^3$ law \cite{shl} (including intermittency corrections),
showing further that the scaling does not necessarily imply the latter.
One of the common assumptions in the description of anomalous transport
is a power law behavior of some input distribution function. For
instance in Langevin or Fokker--Planck approaches, a power law behavior
is generally chosen for the distribution of thermal noise\cite{fog} or
in the spatial correlations\cite{bou}; in deterministic chaotic models,
power--law amplification is used\cite{oku}; in the random walks, long
algebraic tails in step distributions or sticking times are used
\cite{shl}; in fractional diffusion equations, the power of the
fractional derivatives are chosen to describe the anticipated
behavior\cite{zas}. This is not to say that such power--law behavior is
not seen. Indeed experimentally one can justify some of these
assumptions\cite{swi}.

It is instructive to preface our analysis with the results from the
classical diffusion problem in $d-$dimensions in the presence of a
quenched random force $F_\mu (x)$ and thermal activation $\eta _\mu
(t)$. Consider the Langevin equation\cite{bou}
%-----------------------------------------------------------------------
\begin{equation}
\frac{dx_\mu }{dt}=F_\mu (x)+\eta _\mu (t),
\end{equation}
%-----------------------------------------------------------------------
where spatial averages are $\overline{F_\mu (x)}=F_{0,\mu}$,
$\overline{[F_\mu (x)-F_{0,\mu }][F_\nu (x^{\prime })-F_{0,\nu
}]}=G_{\mu \nu }(x-x^{\prime })$ , and the time average of the noise
term is $\overline{\eta _\mu (t)\eta _\nu (t^{\prime })}=2D\delta
(t-t^{\prime })\delta _{\mu \nu }$. One of the crucial issues is the
construction of the statistical correlation function $ G_{\mu \nu }(x)$.
By using a power law dependence for this function, it can be shown that
the dynamics displays diffusion in three dimensions.  Thus, the long
distance correlations, which are characteristic to superdiffusive
dynamics, have to be incorporated into the problem form the very
beginning \cite{bou}.

In this letter we would like to approach the problem from a slightly
different perspective and extend this type of study to the quantum
regime, using the Schr\"odinger equation to describe the dynamics,
instead of the Langevin formulation with the thermal activation. The
fluctuations will emerge from the dynamics of a test particle in the
presence of a correlated chaotic quantum background. We will see that
turbulent--like behavior can be manifest on certain time scales under
fairly general conditions and that power--law assumptions are not
necessary for this.

Since it is known that chaotic dynamics can induce superdiffusive behavior
in classical 1d--maps \cite{oku}, as well as in classical motion coupled to
quantum backgrounds\cite{kus}, a natural starting point is to utilize
random matrix Hamiltonians, which are essentially the quantum counterparts of
classical chaotic systems.
This will provide both a reasonable physical picture as
well as a tractable framework for the analysis of the diffusion process. We
take a model space which is the direct product of the Hilbert space of the
test particle with position $R$, and the finite dimensional (albeit
large) background space defined by a complete basis of states $\left|
i\right\rangle $, with $i=1,...,N\gg 1$. The background, denoted $V_{ij}(R)$
, will be taken to be quenched (time-independent), and chaotic in the sense
that the spatial inhomogeneity is described by a deformed, parametric,
banded random matrix:
%-----------------------------------------------------------------------
\begin{equation}
V_{ij}(R)=U_{0,ij}+U_{1,ij}(R)
\end{equation}
%-----------------------------------------------------------------------
where $U_{1,ij}(R)$ is a real symmetric matrix, and an element of
the Gaussian orthogonal ensemble (GOE)\cite{meh}. As such, it is
characterized by the first two cumulants:
%-----------------------------------------------------------------------
\begin{eqnarray}
\overline{U_{1,ij}(R)} &=&0 \\
\overline{U_{1,ij}(R)U_{1,kl}(R^{\prime })} &=&[\delta
_{ik}\delta _{jl}+\delta _{il}\delta _{jk}]{\cal G}_{ij}(R,R^{\prime }).
\nonumber
\end{eqnarray}
%-----------------------------------------------------------------------
Here the overline indicates the average over the GOE\cite{caio}.
The density of states of the background is defined through the diagonal
matrix $U_{0,ij}=\Omega _i\delta _{ij}$, with the constant average level
density $\rho _0$ ($\rho _0 (\Omega _{i+1}-\Omega _i)=1$).
The correlation function ${\cal G}_{ij}(R,R^{\prime })$ is
parameterized as \cite{wei,pre}
%-----------------------------------------------------------------------
\begin{equation}
{\cal G}_{ij}(R,R^{\prime })=\frac{\Gamma ^{\downarrow }}{2\pi \rho _0 }
\exp \left[ -\frac{(\Omega _i-\Omega _j)^2}{
2\kappa _0^2}\right] G\left ( \frac{R-R^{\prime }}{X_0}\right ).
\end{equation}
%-----------------------------------------------------------------------
This incorporates a bandedness for the random matrix, with an effective width
$N_0=\kappa _0\rho _0$, serving to limit the interaction range to the nearby 
states, an overall strength denoted $\Gamma ^{\downarrow }$ (also known as the
spreading width\cite{pre}), and a spatial correlation function $G(R/X_0)$
with length scale $X_0$, normalized such that $G(0)=1.$  We will assume
that the statistics of the background are translationally invariant,
 ${\cal G}_{ij}(R,R^{\prime })={\cal
G}_{ij}(R-R^{\prime })$, although this is not crucial. This function is
the matrix analog of the correlated noise used in Eq. (2). But instead
of building in long power law tails, we will use $G(x)=\exp [-x^2/2]$,
which provides a rapid spatial decorrelation. Next we couple a test
particle of mass $M$ to this ``chaotic'' background through the
Hamiltonian:
%-----------------------------------------------------------------------
\begin{equation}
H_{ij}(R)=-\delta_{ij} \frac{\hbar ^2}{2M}\partial _R^2+V_{ij}(R).
\end{equation}
%-----------------------------------------------------------------------
Here, as in all formulas we present, $R$ can be interpreted as a variable
of an arbitrary dimensionality, even though some of the formulas we
write explicitly for the 1--d case. Using the Feynman and Vernon
\cite{fey} formalism one can represent the density matrix for our test
particle through the following path integral formula
%-----------------------------------------------------------------------
\begin{eqnarray}
\rho (R,R^\prime ,t) & = & \int d X_0 dY_0
\rho _0(R_0, R_0^\prime )
\int _{R(0)=R_0} ^{R(t)=R} {{\cal D}}R(t)
\int _{R^\prime (0)=R^\prime _0} ^{R^\prime (t)=R^\prime}
{{\cal D}}R^\prime (t) \nonumber \\
 & \times & \exp \left \{ \frac{i}{\hbar }
\left [ S_0(R(t)) - S_0(R^\prime (t)) \right ] \right \}
F(R(t),R^\prime (t),t),
\end{eqnarray}
%-----------------------------------------------------------------------
where $S_0(R(t))$ is just the classical action for a free particle and
$\rho _0(R_0, R_0^\prime )$ is the initial density matrix of the test
particle.  Taking advantage of the large $N_0$--limit, it is possible to
explicitly compute the influence functional for our Hamiltonian in the
adiabatic limit \cite{ann,pre} (which we discuss below), where we find
%-----------------------------------------------------------------------
\begin{equation}
F(R,R^{\prime },t)=\exp \left\{ \frac{\Gamma ^{\downarrow }}\hbar
\int_0^tdt^{\prime }\left[ G([R(t^{\prime })-R^{\prime }(t^{\prime
})]/X_0)-1\right] \right\} .
\end{equation}
%-----------------------------------------------------------------------
Thus the density matrix for the test particle satisfies the following
equation \cite{naka}:
%-----------------------------------------------------------------------
\begin{equation}
 i\hbar\frac{\partial}{\partial t}\rho(R,R',t)=
  \left[  -\frac{\hbar ^2}{2M}(\partial _R^2 -\partial _R^{\prime \;2})
+i\Gamma^\downarrow(G(R,R')-1))\right]\rho(R,R',t).
\end{equation}
%-----------------------------------------------------------------------
In terms of the new variables $r=(R+R^{\prime })/2$ and $s=R-R^{\prime
}$ the solution of this equation can be found through quadratures (easily
verified by direct substitution):
%-----------------------------------------------------------------------
\begin{equation}
\rho (r,s,t)=\int dr^{\prime }\int \frac{dk}{2\pi \hbar }
\rho _0 \left ( r^{\prime },s-\frac{kt}{M}\right )
\exp \left[ \frac{ik(r-r^{\prime })}\hbar +\frac{\Gamma
^{\downarrow }M}{\hbar k}\int_{s-kt/M}^sds^{\prime }[G(s^{\prime
}/X_0)-1]\right]   \label{rho}
\end{equation}
%-----------------------------------------------------------------------
where $\rho_0(r,s)=\rho(r,s,t=0)$ is the initial density matrix at $t=0$.
If we take the initial state to be a Gaussian $\psi _0(R)=
\exp [-R^2/4\sigma ^2]/[2\pi \sigma ^2]^{1/4}$, the initial
density matrix is
%-----------------------------------------------------------------------
\begin{equation}
\rho _0(R,R^{\prime })=\frac 1{\sqrt{2\pi \sigma ^2}}e^{-(R^2+R^{\prime
}{}^2)/4\sigma ^2}=\frac 1{\sqrt{2\pi \sigma ^2}}e^{-(4r^2+s^2)/8\sigma ^2}.
\label{gauss}
\end{equation}
%-----------------------------------------------------------------------
The adiabatic condition in which the influence functional (8) is valid, 
restricts the velocity $V$ of the test particle such that the time scale 
($X_0/V)$ is no greater than that of the background characterized by $\hbar
/\kappa _0$ : $ V_{max}\sim \kappa _0X_0/\hbar $. This can be used to constrain
the average momentum of the initial wavepacket, through the width $\sigma $.
For our initial gaussian, these are related by $\langle P^2\rangle =\hbar
^2/(4\sigma ^2)\sim (MV)^2$, so we require :
%-----------------------------------------------------------------------
\begin{equation}
\sigma \geq \sigma _{\min }=\frac{\hbar ^2}{2MX_0\kappa _0}.
\end{equation}
%-----------------------------------------------------------------------
To extract the diffusive properties of the wavepacket, we compute the
cumulants of the coordinate $R$ directly from the coordinate distribution $
{\cal P}(R,t)=\rho (r,s=0,t)$. This is done by constructing the characteristic
function for the coordinate distribution, $d(k,t)$, defined by taking the
Fourier transform of ${\cal P}(R,t)$:
%-----------------------------------------------------------------------
\begin{eqnarray}
d(k,t) &=&\int dr
\rho _0(r,-kt/M)\exp \left\{ -\frac{ikr}{\hbar } +
\frac{\Gamma
  ^{\downarrow }M}{\hbar k}\left( -\frac{kt}M+\int_{-kt/M}^0dsG(s)\right)
   \right\}  \\
\  &=&\exp \left\{ -\frac 12\left( \frac{\sigma k}{\hbar }\right) ^2-\frac 1
2\left( \frac k{2M\sigma }\right) ^2t^2+\frac{\Gamma ^{\downarrow }MX_0\sqrt{
2}}{k\hbar }\sum_{n=1}^\infty \frac{(-1)^n}{n!(2n+1)}\left( \frac{kt}{\sqrt{2
}MX_0}\right) ^{2n+1}\right\}   \nonumber \\
\  &=&\exp \left[ \sum_{m=1}^\infty \frac{(ik/\hbar)^m}{m!}\left\langle
\; \left\langle R^m\right\rangle \; \right\rangle \right]   \nonumber
\end{eqnarray}
%-----------------------------------------------------------------------
All the cumulants are easily identified.  The second cumulant,
which measures the spreading of the wavepacket, is given by
%-----------------------------------------------------------------------
\begin{eqnarray}
\left\langle \;\left\langle R^2\right\rangle \;\right\rangle  &=&\int
dR\;R^2{\cal P}(R,t)=\left. -\hbar^2\frac{d^2}{dk^2}d(k,t)\right| _{k=0} \\
\  &=&\sigma ^2+\frac{\hbar ^2}{4M^2\sigma ^2}t^2+\frac{\Gamma ^{\downarrow
}\hbar }{3M^2X_0^2}t^3  \label{rr}
\end{eqnarray}
%-----------------------------------------------------------------------
The terms are readily identified. The first is the initial width at
$t=0$, and the second is the natural spreading of Eq. (\ref{gauss}) due
to free expansion, which is the only dynamical contribution when the
background is removed ($\Gamma ^{\downarrow }=0$). The dissipative
contribution which arises from the background displays the diffusion
associated with turbulent backgrounds, namely the $t^3$ character. One
can see that the turbulent--like contribution becomes dominant on the
time scale
%-----------------------------------------------------------------------
\begin{equation}
t_{{\rm T}}\approx \frac{3X_0^2\hbar }{4\sigma ^2\Gamma ^{\downarrow }}\ .
\end{equation}
The momentum distribution ${\cal P}(P,t)$ and its characteristic function
$D(s,t)$, are given by
%-----------------------------------------------------------------------
\begin{eqnarray}
{\cal P}(P,t) &=&\int dRdR^{\prime }\exp \left (\frac{iP(R-R^{\prime
})}{\hbar }\right )\rho
(R,R^{\prime },t)
  \\ &=&
\int ds\;\exp \left (\frac{iPs}{\hbar}\right ) D(s,t) , \\
\left\langle \;\left\langle P^2\right\rangle \;\right\rangle  &=&-\left. \hbar
^2\frac{d^2}{ds^2}D(s,t)\right| _{s=0}=\frac{\hbar ^2}{4\sigma ^2}+\frac{
\hbar \Gamma ^{\downarrow }}{X_0^2}t.\label{rhop}
\end{eqnarray}
%-----------------------------------------------------------------------
One can see from Eq. (\ref{rhop}) that in the absence of  coupling to the
background $ (\Gamma ^{\downarrow }=0)$, the momentum cumulant is
constant and given by the usual value for a wavepacket. The coupling to
the background makes the momentum variance increase linearly with time.
Because this turbulent--like behavior is limited to the adiabatic
regime, the maximum time scale for turbulent--like diffusion to be
present is given by the condition $\left\langle \;\left\langle
P^2\right\rangle \;\right\rangle ^{1/2}\sim MV_{\max }\sim MX_0\kappa
_0/\hbar$, or:
%-----------------------------------------------------------------------
\begin{equation}
t_{\max }\approx \frac 13t_{{\rm T}}\left[ \left( \frac \sigma {\sigma
_{\min }}\right) ^2-1\right]
\end{equation}
%-----------------------------------------------------------------------
which depends only upon the initial width of the wavepacket $\sigma $.
Hence for times on the scale $t_{{\rm T}}\leq t\leq t_{\max }$, the
diffusion of the wavepacket will have a turbulent--like character.
($t_{\max}>t_{\rm T}$ requires only that $\sigma>2\sigma_{\min}$). For
$t\geq t_{\max }$, the character of the interaction with the background
changes over to a diabatic behavior, where the above form for the
influence functional is no longer valid. This is not to say however that
the dynamics ceases to have a turbulent--like behavior, only that our
adiabatic expression for the influence functional (8) has a limited range
of applicability.

Similar to the Langevin approach in (2), the anomalous diffusion arises
from the properties of the spatial correlations, but for quite different
reasons.  Because we are using a random matrix ensemble to model the
background properties,  the correlation function $G(x)$ must be positive
definite, and can be classified by its short distance behavior:
$G(x)\approx 1-c|x|^\alpha +\cdots $ where the range is restricted to
$0\leq \alpha \leq 2$ and $x=R/X_0 $\cite{caio}. If we want a smoothly
correlated background then $\alpha =2$, while for a Brownian motion type
spatial fluctuations $\alpha =1$.  We will only consider here the case
of smooth spatial correlations,  $\alpha =2$. (The case $\alpha
<2$ would be interesting to consider further, as the diffusion would be
characterized by very long spatial tails and infinite moments, analogous to the
Levy stable laws\cite{bou}.)  The Gaussian
correlation function provides {\it generic} results for any $\alpha =2$
correlator, which can be seen from the definition of the second cumulant. 
As the inverse is also a Gaussian, neither the spatial
correlation nor its Fourier transform exhibit any long range
correlations. All long range diffusive behavior emerges from spatial
inhomogeneities on the scale $X_0$. If we vary this scale, taking the
limit $ X_0\rightarrow 0$, the anomalous diffusion is enhanced, as can
be seen  directly from $\left\langle \;\left\langle R^2\right\rangle \;
\right\rangle$. The opposite limit, $X_0\rightarrow \infty $,
corresponds to a constant random background ($U_{1,ij}(R)$ is replaced
with a fixed random matrix), and the turbulent diffusion vanishes, since
the spatial domain  on which it is active is never reached.

By examining the dynamics of a wave packet in a chaotic background, we have
found that turbulent--like  diffusion can emerge under very general
circumstances,  with only the input of the short distance spatial correlations
in the background on a finite scale $X_0$:  no power law distributions are
assumed.  Further, our results for the full coordinate and momentum
distributions,  ${\cal P}(R,t)$ and ${\cal P}(P,t)$, do not exhibit the usual
scaling behaviors or power law properties used in previous studies. Our
dynamics is a statistical limit which emerges from the random matrix solution
to the influence functional, and as the fluctuations are gaussian, cumulants 
higher than second order are not invoked. As with classical turbulent
diffusion, which can be generalized to included intermittency corrections and
so forth, a more general class of this turbulent-like quantum diffusion can be
explored by considering various types of backgrounds. This would include, for
example, stochastic rather smooth  spatial correlation functions $G(x)$
characterized by $\alpha<2$, corrections to the density of states for
non-constant behaviors, or inclusion of higher cumulants in the background. In
addition, the role of $\hbar$, in particular, how the turbulent
diffusion survives the limit $\hbar \rightarrow 0$. These might provide a more
general formulation of the quantum analog of diffusion in chaotic backgrounds,
in which a classical limit might eventually recover intermittency corrections.


\begin{references}

\bibitem{bou}  J. Bouchaud and A. Georges, {\sl Phys. Rep.} {\bf 195}
(1990) 127; M. Isichenko, {\sl Rev. Mod. Phys.} {\bf 64} (1992) 961.

\bibitem{long}  See for example: {\sl L\'evy flights and related topics
in physics}, Eds. M.F. Shlesinger, G.M. Zaslavsky and U. Frisch ,
(Springer-Verlag, New York 1995).

\bibitem{ric}  L. Richardson,{\sl Proc. R. Soc. (London)} {\bf A110}
(1926) 709.

\bibitem{kol}  A. Kolmogorov, {\sl J. Fluid. Mech.} {\bf 13} (1962) 82.

\bibitem{Man}  B. Mandelbrot, {\sl J. Fluid Mech. }{\bf 62} (1974) 331;
U. Frisch, P. Sulem, M. Nelkin, {\sl J. Fluid Mech. }{\bf 87} (1978)
719.

\bibitem{shl}  M. Shlesinger, B. West and J. Klafter, {\sl Phys. Rev.
Lett.}  {\bf 58} (1987) 1100; {\sl Phys. Scr. }{\bf 40} (1986) 445.

\bibitem{fog}  H. Fogedby, {\sl Phys. Rev. Lett.} {\bf 73} (1994) 2517.

\bibitem{oku}  A. Okubo, V. Andeasen and J. Mitchell, {\sl Phys. Lett.}
{\bf  A105} (1984) 169; T. Geisel and S. Thomae, {\sl Phys. Rev. Lett.}
{\bf 52} (1984) 1936.

\bibitem{zas}  G. Zaslavsky, M. Edelman and B. Niyazov, Courant
Institute preprint (1996).

\bibitem{swi}  T. Solomon, E. Weeks and H. Swinney, {\sl Phys. Rev.
Lett.}  {\bf 71} (1993) 3975.

\bibitem{kus}  D. Kusnezov, {\sl Phys. Rev. Lett.} {\bf 72} (1994) 1990.

\bibitem{meh}  M. Mehta, {\sl Random Matrices} (Academic Press, San
Diego, 1991).

\bibitem{caio}  D. Kusnezov and C. Lewenkopf, {\sl Phys. Rev.} {\bf E53}
(1996) 2283.

\bibitem{wei}  D. Brink, J. Neto and H. Weidenmuller, {\sl Phys. Lett.}
{\bf  80B} (1979) 170.

\bibitem{pre}  A. Bulgac, G. Do Dang and D. Kusnezov, {\sl Phys. Rev.}
{\bf  E54} (1996) 3468.

\bibitem{fey}  R. Feynman and F. Vernon, {\sl Ann. Phys. (NY)} {\bf 24}
(1963) 118.

\bibitem{ann}  A. Bulgac, G. Do Dang and D. Kusnezov, {\sl Ann. Phys.
(NY)}  {\bf 242} (1995) 1.

\bibitem{naka}  A. Bulgac, G. Do Dang and D. Kusnezov, {\sl Chaos,
Solitons and Fractals} (to appear, July 1997) and xxx.lanl.gov e--Print
archive nucl-th/9612011.


\end{references}
\end{document}